\documentclass{article}
\usepackage{spconf}

\usepackage{diagbox} 
\usepackage{graphicx}
\usepackage{epstopdf}
\usepackage{psfrag}
\usepackage{subfigure}
\usepackage{url}
\usepackage{stfloats}
\usepackage{amsfonts,amssymb,amsmath,bm,paralist,theorem,cite,ifthen,color,nccmath}
\usepackage{caption}
\usepackage{calc}
\usepackage{enumerate}
\usepackage{multirow}
\usepackage{makecell}
\usepackage[linesnumbered,ruled]{algorithm2e}
\usepackage{setspace}
\usepackage{array}
\usepackage{float}
\usepackage{soul}

\usepackage{lipsum} 

\usepackage[colorlinks,
            linkcolor=blue,
            anchorcolor=red,
            citecolor=red]{hyperref}
\def\T{\mathrm T}
\def\H{\mathrm H}

\newcommand\Nr{\ensuremath{ N_{\rm r} }}

\newcommand\Nw{\ensuremath{ N_{\rm w} }}
\newcommand\Nu{\ensuremath{ N_{\rm u} }}

\newcommand\rmj{\ensuremath{ \mathrm j }}

\newcommand\rmc{\ensuremath{ \mathrm{c} }}
\newcommand\rms{\ensuremath{ \mathrm{s} }}

\newcommand\rmtr{\ensuremath{ \mathrm{Tr} }}
\newcommand\rmt{\ensuremath{ \mathrm{t} }}

 \usepackage[all=normal,paragraphs=normal,floats=normal,wordspacing=tight,charwidths=tight,mathdisplays=normal,leading=normal]{savetrees}
\title{Tri-Hybrid Beamforming Design for \\Integrated Sensing and Communications}
\name{Tianyu Fang$^*$, Mengyuan Ma$^*$, Markku Juntti$^*$, Nhan Thanh Nguyen$^*$}
\address{$^*$Centre for Wireless Communications, University of Oulu, P.O.Box 4500, FI-90014, Finland}
\begin{document}
\setlength{\abovedisplayskip}{3.0pt}
\setlength{\belowdisplayskip}{3.0pt}

%
\maketitle
\begin{abstract}
Tri-hybrid beamforming architectures have been proposed to enable energy-efficient communications systems in extra-large-scale antenna arrays using low-cost programmable metasurface antennas. We study the tri-hybrid beamforming design for integrated sensing and communications (ISAC) to improve both communications and sensing performances. Specifically, we formulate a multi-objective optimization problem that balances communications signal-to-noise ratio (SNR) and the sensing power at a target direction, subject to constraints on the total power consumption and physical limitations inherent to the tri-hybrid beamforming architecture. We develop an efficient iterative algorithm in which the variables are updated in a closed form at each iteration, leading to a low-complexity and fast-execution design. Numerical results show that the tri-hybrid architecture improves spatial gain and energy efficiency, though with reduced beam alignment capability compared to conventional hybrid beamforming architectures. 

\end{abstract}
\vspace{-0.25cm}
\section{Introduction}\vspace{-0.25cm}
\label{sec:intro}
Millimeter-wave (mmWave) and terahertz (THz) communications enable integrated sensing and communications (ISAC) with extra-large-scale antenna arrays, supporting ultra-high data rates and fine sensing resolution~\cite{Lu2024}. However, conventional digital beamforming requires a dedicated radio frequency (RF) chain per antenna, which can incur prohibitively high power consumption and hardware cost, posing significant challenges for ISAC efficiency ~\cite{jiang2024THz}. The recently introduced tri-hybrid beamforming (tri-HBF) architecture integrates HBF with programmable metasurface antennas, enabling the use of more antenna elements while substantially reducing the number of required RF chains~\cite{castellanos2025embracing}. Consequently, tri-HBF is a promising solution to mitigate the power consumption bottleneck in mmWave/THz ISAC systems.

Compared with conventional HBF, tri-HBF~\cite{castellanos2025embracing} introduces reconfigurable antennas as a third beamforming layer, enabling the compact integration of a large number of reconfigurable elements. In this design, the beamforming is performed across three domains: digital baseband, analog network, and electromagnetic reconfigurable structures, realized through dynamic metasurface antennas (DMA)~\cite{Nir2021DMA} or reconfigurable intelligent surfaces (RIS)~\cite{basar2024reconfigurable,an2023SIM}. 
Recent works~\cite{gavras2023full,Bayraktar2024near,zhu2025the,alexandropoulos2025extremely} have investigated DMA-based ISAC transceivers and demonstrated their potential benefits. Building on this direction, the tri-HBF architecture introduces a three-stage beamforming structure that can further reduce the number of phase shifters compared to HBF solutions while maintaining high spatial multiplexing gains. Therefore, tri-HBF can outperform the conventional digital and HBF architecture in energy efficiency (EE) \cite{castellanos2023energy, heath2025tri, castellanos2025embracing}. However, the tri-HBF architecture with DMA for ISAC systems remains unexplored, as most existing works focus on deriving feasible solutions rather than addressing the joint design and optimization of tri-HBF beamforming \cite{castellanos2023energy, heath2025tri, castellanos2025embracing}. We are motivated to bridge this gap in this work.


We consider a mmWave ISAC system with a large antenna array, implemented using a tri-HBF architecture. We aim to enhance both communications signal-to-noise ratio (SNR) and the transmit power toward a direction by jointly optimizing the digital, analog, and DMA beamformers. The formulated problem is non-convex and highly challenging because of the constant-modulus constraint on the analog precoding coefficients (as in the conventional HBF designs) and the additional hardware feasibility constraints imposed by the DMA. To overcome these challenges, we develop an efficient iterative algorithm that exploits the problem structure to enable closed-form updates of the beamformers in each iteration, leading to a low-complexity and fast-execution
design. Simulation results demonstrate that the tri-hybrid architecture significantly enhances the spatial gain and EE of the communications functionality at a marginal performance degradation of beam alignment compared to the conventional HBF schemes. To the best of the authors’ knowledge, this is among the first works to optimize tri-hybrid beamformers for ISAC systems.

\vspace{-0.25cm}
\section{System model and problem formulation}\vspace{-0.25cm}
\label{sec:system model}
\begin{figure}[t]
\center
\includegraphics[width=0.45\textwidth]{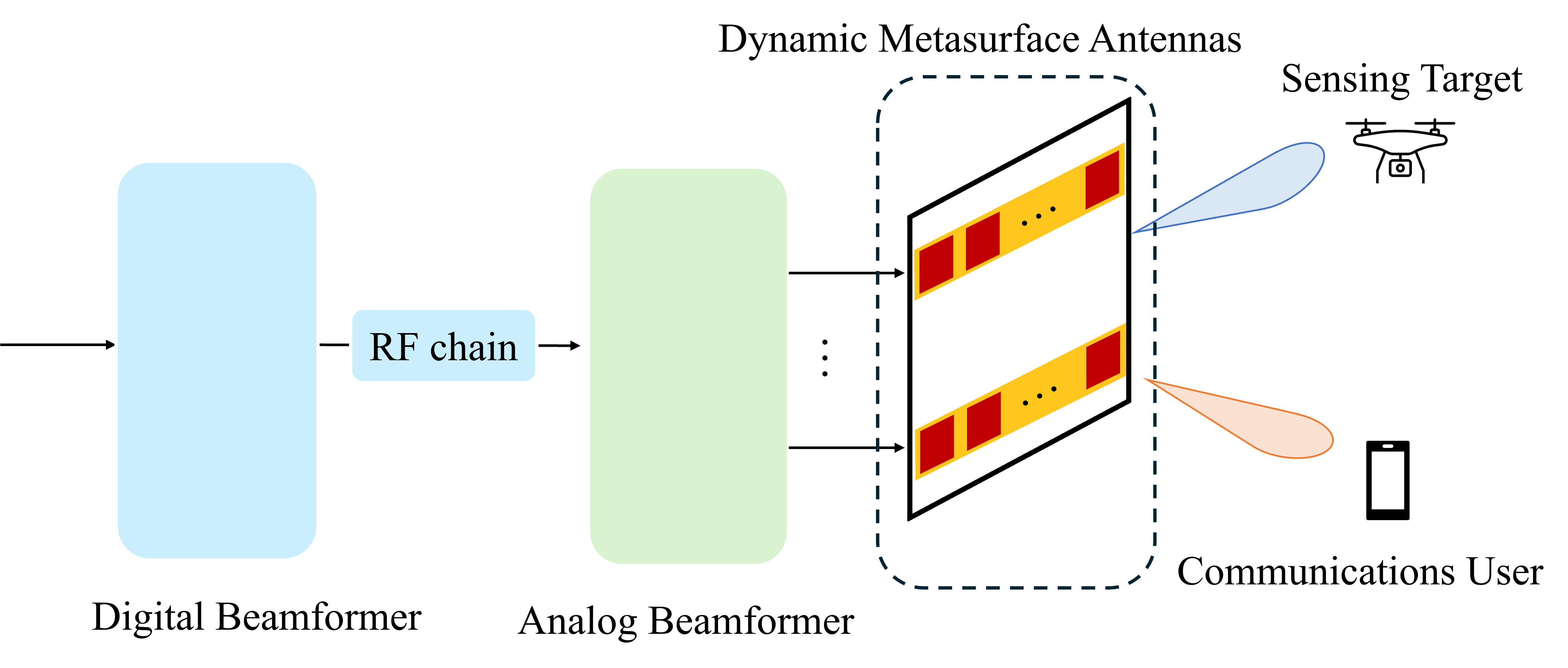}
    \caption{Illustration of the system model.}
    \label{fig:system model}
\end{figure}

We consider a mmWave mono-static ISAC system, where a base station (BS) employs a tri-HBF architecture with the DMA as illustrated in Fig.~\ref{fig:system model}.
The BS transmits data signals to one single-antenna communications user and simultaneously leverage the signal to probe a specified direction of sensing target. For simplicity, we assume that the BS employs only one RF chain to serve the single data stream toward the user. 

\textbf{Tri-HBF Structure:} Let $w\in\mathbb C$ and $\mathbf f=[f_1,\ldots,f_{\Nw}]^\T\in\mathbb C^{\Nw\times 1}$ be the digital and  analog beamformers at the BS, repectively, where $\Nw$ is the number of waveguides in the DMA. The analog beamforming coefficients are subject to the unit-modulus constraints:
\begin{equation}\label{analog constraint}
    |f_i|=1, \quad \forall i\in\mathcal \Nw,
\end{equation}
where $\mathcal\Nw\triangleq \{1,\ldots,\Nw\}$.
Each waveguide of DMA contains $\Nu$ reconfigurable radiating elements. Thus, there are $\Nr=\Nw\Nu$ radiating elements in total, indexed by $\mathcal N_{\rm r}=\{1,\ldots,\Nr\}$. Each radiating element captures the signal from its corresponding waveguide and applies a tunable phase shift before transmission. 

Let $\mathbf m=[m_{1},\ldots, m_{\Nr}]^\T\in\mathbb C^{\Nr}$ be the vector of the coefficients associated with all the radiating elements, where \cite{chen2025energy}
\begin{equation}\label{DMA constraint1}
    m_j=q_j\alpha_j, \quad \forall j\in\mathcal\Nr.
\end{equation}
Here, $q_{j}\in\mathbb C$ represents the propagation gain from the input port of waveguide to the radiator. This coefficient depends on the waveguide material properties and can be measured in advance~\cite{heath2025tri}, therefore we assume $q_{j}$ is known at the BS. Moreover, $\alpha_{j}\in\mathbb C$ is the tunable coefficient applied by the $j$-th radiating element, which satisfies the Lorentzian constraint \cite{nir2019dynamic}:
\begin{align}\label{DMA constraint2}
    \alpha_{j}={(\mathrm j+\psi_{j})}/{2},\ \text{where~} |\psi_{j}|=1, \forall j\in\mathcal \Nr.
\end{align}
Since the $\Nu$ radiating elements in a waveguide share a single input port, the $\Nr$ antenna elements are naturally partitioned into $\Nw$ subarrays, resulting in a block-diagonal structure of the DMA beamformer, denoted by~\cite{castellanos2025embracing} 
\begin{equation}\label{DMA constraint3}
   \! \mathbf M\!=\!\mathrm{blkdiag}\left(\mathbf m_1,\ldots,\mathbf m_{\Nw}\right),
\end{equation}
where $\mathbf m_i\triangleq [m_{(i-1)\Nu+1},\ldots,m_{i\Nu}]^\T\in\mathbb C^{\Nu}, \forall i\in\mathcal\Nw$. 


{\bf Communications Model:} Let $s$ denote the transmit symbol, with $\mathbb E \left[|s|^2\right]=1$. The received signal at the communications user is modeled as
\begin{align} \label{eq_comm_model}
    y_\rmc=\bar{\mathbf h}^\H\mathbf M\mathbf fws+n,
\end{align}
where $\bar{\mathbf h}$ denotes the channel vector between the BS and the user, assumed to be known at the BS in our initial analysis herein (although we admit that its practical estimation is an important problem on its own), and $n\sim \mathcal{CN}(0, \sigma^2)$ represents the additive white Gaussian noise with variance $\sigma^2$.The SNR for communications user is given by
\begin{align}\label{eq_SNR}
    \text{SNR} = |\mathbf h^\H\mathbf M\mathbf fw|^2,
\end{align}
where $\mathbf h\triangleq \bar{\mathbf h}/\sigma$ denotes normalized channel vector.

{\bf Radar Model:} Let $(\theta,\phi)$ denote the azimuth and elevation angles specifying the direction of the sensing target relative to the BS’s DMA.  
Denote by $\mathbf{g}(\theta,\phi)$ the corresponding steering vector at the BS.  
For ease of exposition, we omit the explicit dependence on $(\theta,\phi)$ in the following.  
The radar probing signal toward the target direction is expressed as  
\begin{equation}
    y_{\mathrm{s}} = \mathbf{g}^\H \mathbf{M} \mathbf{f} \, w s.
\end{equation}
The transmit power for sensing is computed as~\cite{liao2024power}:
\begin{align}\label{eq_sensing_power}
    P_{\text{sensing}} = \mathbb{E}\left[ |y_{\mathrm{s}}|^2 \right] = \|\mathbf g^\H \mathbf M \mathbf f w\|^2.
\end{align}

{\bf Problem Formulation:} We jointly design the DMA beamforming matrix $\mathbf{M}$, the analog beamforming vector $\mathbf{f}$, and the digital beamforming weight $w$ in the tri-HBF architecture to achieve high communications SNR and radar sensing power. The problem of interest is formulated as
\begin{subequations}\label{P1}
    \begin{align}
    \max_{ w,\mathbf f,\mathbf M}\,\, &\delta_\rmc |\mathbf h^\H\mathbf M\mathbf fw|^2+\delta_\rms |\mathbf g^\H\mathbf M\mathbf fw|^2\\
    \label{P1C1}\text{s.t.}\,\, &\eqref{analog constraint},\eqref{DMA constraint1},\eqref{DMA constraint2},\eqref{DMA constraint3}\\
    \label{P1C4}&\|\mathbf M\mathbf f w\|^2\leq P_\rmt,
\end{align}
\end{subequations}
where weights $ \delta_\rmc \geq 0$ and $\delta_\rms\geq 0$ control the trade-off between communications and sensing performance, while $P_\rmt$ denotes the total transmit power budget. Constraints in \eqref{P1C1} represent the hardware limitations of analog and DMA beamformers while constraint \eqref{P1C4} enforces the antenna power constraint. For the design objective in \eqref{P1}, the fully-digital beamforming admits a closed-form solution. In contrast, problem~\eqref{P1} is inherently non-convex and NP-hard due to the unit-modulus constraint and the strong coupling among the optimization variables. We propose an efficient iterative algorithm that decouples the variables and derives closed-form solutions for them, as detailed below.

\vspace{-0.25cm}
\section{Tri-hybrid Beamforming Design}\vspace{-0.25cm}
\label{sec:method}
We employ alternating optimization as the general framework to decouple the design vaiables. Sepecifically, for given $\mathbf f$ and $\mathbf M$, we first obtain a closed-form solution to $w$. Then, the problem can be simplified into a single-ratio fractional programming, which is solved by using Dinkelbach’s transform~\cite{dinkelbach1967nonlinear}. The whole procedure is elaborated in the following. 

{\bf Problem Reformulation:}
For any given analog and DMA beamformer $\mathbf f$ and $\mathbf M$, problem \eqref{P1} is reduced to
\begin{equation*}
     \max_{ w}\,|w|^2 \!\left(\delta_\rmc |\mathbf h^\H\mathbf M\mathbf f|^2\!+\!\delta_\rms |\mathbf g^\H\mathbf M\mathbf f|^2\right), \ \text{s.t.}\,|w|^2\leq \frac{P_\rmt}{\|\mathbf M\mathbf f\|^2},
\end{equation*}
   whose optimal solution is any $w$ that satisfies
   \begin{align}\label{eq_sol_w}
       |w|^2=\frac{P_\rmt}{\|\mathbf M\mathbf f\|^2}.
   \end{align}
   By substituting it back into problem \eqref{P1}, we obtain the following equivalent problem:
\begin{equation}\label{P2}
     \max_{ \mathbf f,\mathbf M}\,\frac{\delta_\rmc |\mathbf h^\H\mathbf M\mathbf f|^2\!+\!\delta_\rms |\mathbf g^\H\mathbf M\mathbf f|^2}{\rmtr(\mathbf M\mathbf M^\H)},\, \,
  \text{s.t.}\,\, \eqref{P1C1},
\end{equation}
where we have used the fact that
\begin{equation}
    \|\mathbf M \mathbf f\|^2= \sum_{i=1}^{\Nw} \sum_{j=1}^{\Nu}
   \bigl| m_{(i-1)\Nu + j} f_i \bigr|^2
 = \rmtr \!\left( \mathbf M \mathbf M^{\mathrm H} \right).
\end{equation}
Observing that problem \eqref{P2} is a single-ratio fractional program, we adopt the Dinkebach's transform to recast it as
\begin{equation}\label{P3}
    \max_{ \mathbf f,\mathbf M} {\delta_\rmc |\mathbf h^\H\mathbf M\mathbf f|^2\!+\!\delta_\rms |\mathbf g^\H\mathbf M\mathbf f|^2}\!-\!z{\rmtr(\mathbf M\mathbf M^\H)},\
   \text{s.t.}\,\, \eqref{P1C1},
\end{equation}
where $z$ is an introduced auxiliary variable. Problem~\eqref{P3} can be solved by iteratively updating $z$, $\mathbf f$ and $\mathbf M$. Specifically, in iteration $t+1$, $z$ admits the following closed-form solution:
\begin{align}\label{update for z}
    z^{[t+1]}\!=\frac{(\delta_\rmc |\mathbf h^\H\mathbf M^{[t]}\mathbf f^{[t]}|^2\!+\!\delta_\rms |\mathbf g^\H\mathbf M^{[t]}\mathbf f^{[t]}|^2)}{\rmtr(\mathbf M^{[t]}\mathbf M^{{[t]}^\H})},
\end{align}
We will derive the updates for $\mathbf M$ and $\mathbf f$ with $z$ fixed.

{\bf DMA and Analog Beamformer Updates:} We first reformulate problem \eqref{P2} into a more compact form. Specifically, we collect  ${q_{j}}$ and ${\psi_{j}}$ into vectors $\mathbf q$ and $\bm \psi$, respectively, such that $ \mathbf m=(\rmj \mathbf q+\bm\psi\circ \mathbf q)/2$, where $\circ$ denotes the element-wise product. With $\mathbf H\triangleq \mathrm{blkdiag}\left(\mathbf h(1:\Nu),\ldots, \mathbf h((\Nw-1)\Nu+1:\Nw\Nu) \right)$, we obtain
\begin{align*}
    |\mathbf h^\H\mathbf M\mathbf f|^2=|\mathbf m^\T\mathbf H^*\mathbf f|^2=|\mathbf m^\H\mathbf H\mathbf f^*|^2=\mathbf m^\H\mathbf H\mathbf f^*\,\mathbf f^\T\mathbf H^\H\mathbf m,
\end{align*}
where $(\cdot)^*$ denotes the element-wise complex conjugation.
Using this result and with $\mathbf f$ fixed, problem \eqref{P3} reduces to
\begin{subequations}\label{P4}
    \begin{align}
    \max_{ \mathbf m}\,\, &\mathbf m^\H\mathbf A\mathbf m-z\mathbf m^\H\mathbf m\\
 \label{P4C1}   \text{s.t.}\,\, &\mathbf m=(\rmj \mathbf q+\bm\psi\circ \mathbf q)/2, \\
\label{P4C2}    &|[\bm\psi]_m|=1, \forall m\in\mathcal N_{\rm r},
\end{align}
\end{subequations}
where we have denoted $\mathbf A\triangleq \delta_\rmc\mathbf H\mathbf f^*\,\mathbf f^\T\mathbf H^\H+\delta_\rms\mathbf G\mathbf f^*\,\mathbf f^\T\mathbf G^\H$, and $\mathbf G$ is constructed analogously to $\mathbf H$.


To solve \eqref{P4}, we add $z\bm\psi^\H\bm\psi $ to the objective function of \eqref{P4}, considering that $\bm\psi^\H\bm\psi=\Nr$ is satisfied at any feasible point. With such a transformation, the objective function becomes convex without loss of optimality. Define $f(\bm\psi)=\mathbf m^\H\mathbf A\mathbf m-z\mathbf m^\H\mathbf m+z\bm\psi^\H\bm\psi$. With $\mathbf Q\triangleq \mathrm {diag}(\mathbf q)$, the gradient vector and Hessian matrix of $f(\bm\psi)$ are expressed as 
\begin{align}
\label{gradient of f}&\nabla_{\bm\psi} f(\bm\psi)=(2\mathbf A\mathbf m-2z\mathbf m)\circ\mathbf q^*+2z\bm\psi,\\
    &\nabla^2_{\bm\psi} f(\bm\psi)=\mathbf Q^\H\mathbf A\mathbf Q-z\mathbf Q^\H\mathbf Q+z\mathbf I.
\end{align}
 Since $\mathbf A $ is a semi-positive definite matrix and $0<|q_{m,n}|<1$, the Hessian matrix must be positive definite, justifying the convexity of $f(\bm\psi)$ with respect to $\bm\psi$. 
Based on the first-order Taylor approximation \cite{boyd2004convex}, problem \eqref{P4}  can be approximated at iteration $t$ as
\begin{equation}\label{P5} 
    \max_{ \bm\psi}\,\, \Re\left\{\bm\psi^\H  \nabla_{\bm\psi} f(\bm\psi)\big|_{\bm\psi=\bm\psi^{[t]}} \right\},\,\,
   \text{s.t.}\,\, 
    \eqref{P4C2},
\end{equation}
whose optimal solution is given by 
\begin{align}\label{update for M}
    \bm\psi^{[t+1]}=\exp\left(\mathrm j \angle\left(\nabla_{\bm\psi} f(\bm\psi)\big|_{\bm\psi=\bm\psi^{[t]}}\right)\right),
\end{align}
where $\angle(\mathbf{x})$ returns the phases of all the elements of vector $\mathbf{x}$. Then, the DMA beamformer can be obtained based on \eqref{P4C1}.

With the other variables fixed, the subproblem with respect to $\mathbf f$ is written as
\begin{subequations}\label{P5}
    \begin{align}
    \max_{ \mathbf f}\,\, & \mathbf f^\H\mathbf M^\H(\delta_\rmc\mathbf h\mathbf h^\H+\delta_\rms\mathbf g\mathbf g^\H)\mathbf M\mathbf f,\,\, \text{s.t.}\,\, \eqref{analog constraint}.
\end{align}
\end{subequations}
By a similar approach as above, we can obtain $\mathbf{f}$ as
\begin{align}\label{update for f}
   \hspace{-0.2cm} \mathbf f^{[t+1]}\!=\!\exp\!\left(\!\mathrm j \angle\!\left(\!\mathbf M^{[t+1]^\H}\!(\delta_\rmc\mathbf h\mathbf h^\H\!+\!\delta_\rms\mathbf g\mathbf g^\H)\mathbf M^{[t+1]}\mathbf f^{[t]}  \!\right)\!\!\right).
\end{align}

\textbf{Overall Design:} The proposed low-complexity tri-HBF design is summarized in Algorithm~\ref{Alg: DT-SGPI}. Specifically, given a feasible initial point $\{\mathbf M^{[0]},\mathbf f^{[0]}\}$, the auxiliary variable $z$, the DMA beamformer $\mathbf M$, and the analog beamformer $\mathbf f$ are iteratively updated in each iteration until the objective value in \eqref{P1} converges, as seen in steps 3, 4, and 5, respectively. The objective value is non-decreasing after each update and bounded on the feasible set, ensuring monotone convergence. This will be numerically confirmed in the simulation results. The complexity of Algorithm \ref{Alg: DT-SGPI} is dominated by the update of the DMA beamformer $\mathbf m$ in \eqref{update for M}. This requires a complexity of $\mathcal O(\Nr^2)$. Thus, the overall complexity of  Algorithm \ref{Alg: DT-SGPI} is $\mathcal O(I\Nr^2)$, where $I$ is the number of iterations.

\begin{algorithm}[t]
    \small
	\textbf{Initialize}: $t\leftarrow0$, feasible point $\{\mathbf M^{[0]},\mathbf f^{[0]}\}$\;
	\Repeat{convergence of objective value in \eqref{P1}}{
		
        Update $z^{[t+1]}$ using \eqref{update for z}\;

        Update $\mathbf{m}^{[t+1]}$ using \eqref{update for M} and \eqref{P4C1}\;
        Update $\mathbf{f}^{[t+1]}$ using \eqref{update for f}\;
 	$t\leftarrow t+1$\;
	}	
     Scale $w$ to satisfy the power constraint \eqref{P1C4}.
	\caption{Algorithm for Solving Problem~\eqref{P1}}
	\label{Alg: DT-SGPI}				
\end{algorithm}

\section{Simulation Results}
\label{sec:illust}

\textbf{Simulation Setup:} In this section, we numerically evaluate the proposed design. We set  $\Nu=16, \Nw=8, P_\rmt=10$ dBm and the noise variance is $\sigma^2=0$ dBm. We consider a DMA-based BS operating at a carrier frequency of $f_c=28$ GHz corresponding to a wavelength of $\lambda_c=1.07$ cm. The inter-waveguide spacing is set to $\lambda_c/2$ while the spacing between adjacent rading elements is $\lambda_c/5$ \cite{zhang2022beam}. 
Following~\cite{chen2025energy}, the propagation gain is modeled as $q_j = e^{-d_j(\nu + \mathrm{j}\varpi)}$, where $d_j$ denotes the physical distance between the input and the element, $\nu = 0.6\,\mathrm{m}^{-1}$ is the attenuation coefficient, and $\varpi = 827.67\,\mathrm{m}^{-1}$ is the wavenumber. The channel vector is modeled using the Saleh--Valenzuela scattering model~\cite{ayach2014spatially}. The $\Nr$ DMA components are arranged in a rectangular shape, similar to a uniform planar array (UPA). Thus, the DMA steering vectors are identically modeled as those for UPAs \cite{fang2025optimal}.
The azimuth angle $\theta$ is independently drawn from the uniform distribution $\mathcal U(-\pi/3,\pi/3)$, while the elevation angle $\phi$ is sampled from $\mathcal U(\pi/6,5\pi/6)$. All the simulation results are averaged over 100 independent channel realizations.

\begin{figure}[htb]

\begin{minipage}[b]{1\linewidth}
  \centering
  \centerline{\includegraphics[width=0.8\linewidth]{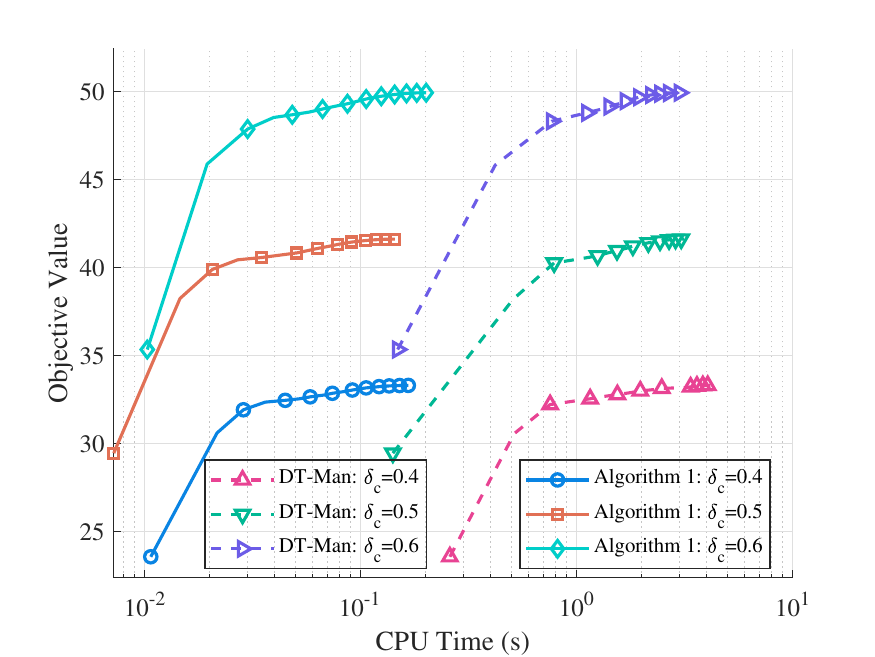}}
\end{minipage}

\caption{\small Convergence behavior of Algorithm \ref{Alg: DT-SGPI}.}
\label{fig:convergence}
\end{figure}

Fig.~\ \ref{fig:convergence} illustrates the convergence behavior as well as the required run time until convergence for Algorithm \ref{Alg: DT-SGPI} under different weight coefficients. For comparison, we include the baseline ``DT-Man”, wherein subproblems \eqref{P4} and \eqref{P5} are solved using a manifold-based conjugate gradient method~\cite{yu2016AMmanifold}. We show the convergences for different values of $\delta_\rmc$ with $\delta_\rms = 1 - \delta_\rmc$. Note that in this figure, each marker corresponds to every second optimization iteration. The objective value is plotted starting from the first iteration, and the initial value is omitted. As observed, the proposed algorithm consistently achieves a monotonic increase in the objective value and converges rapidly  and stably. 
Moreover, Algorithm~\ref{Alg: DT-SGPI} requires a comparable number of iterations as the DT-Man scheme but converges significantly faster in terms of CPU time. This is because DT-Man  typically performs line search to determine a suitable step size, while our proposed method enable updating beamformers with closed-form solutions. 

\begin{figure}[t!]
    \hspace{-0.3cm}
    \subfigure[Achievable rate vs. sensing power.]{%
        \includegraphics[width=0.52\linewidth]{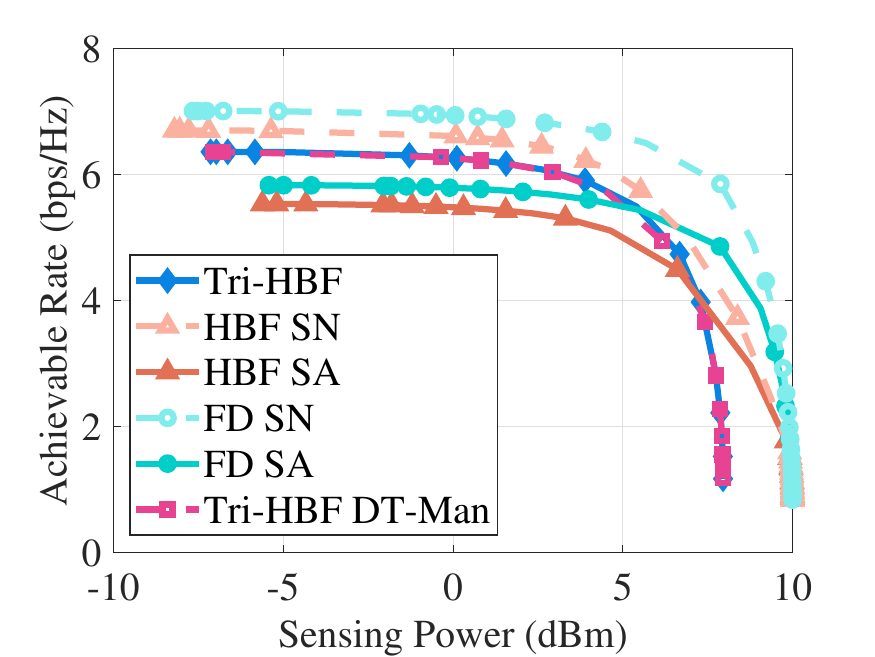}%
        \label{fig:AR_vs_SP}
    }
    \hspace{-0.4cm}
    \subfigure[EE vs. sensing power.]{%
        \includegraphics[width=0.52\linewidth]{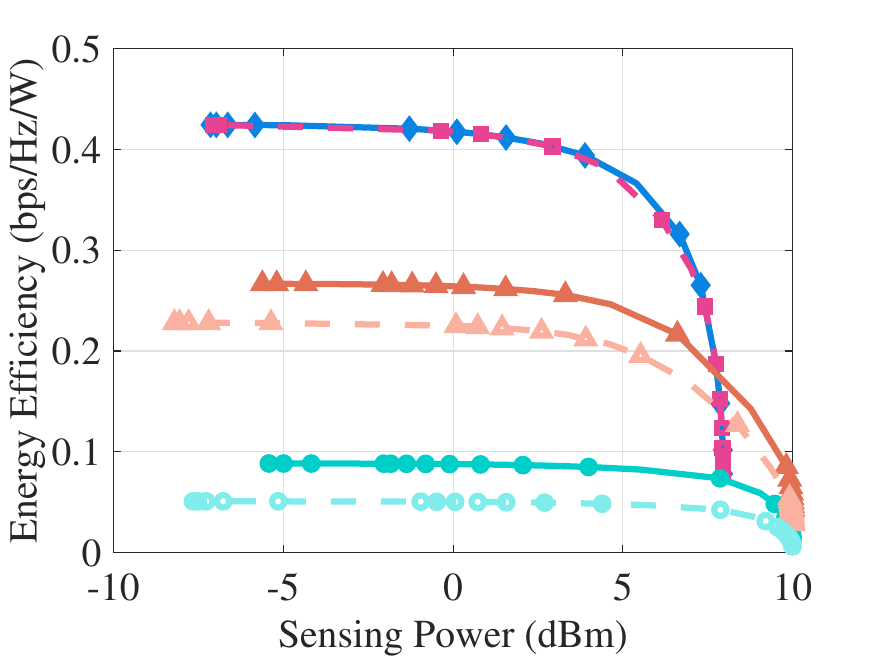}%
        \label{fig:EE_vs_SP}
    }
    \vspace{-3mm}
    \caption{\small Performance tradeoff of considered architectures.}
    \label{fig:SR_EE_tradeoff}
\end{figure}
Fig.~\ref{fig:SR_EE_tradeoff} shows the tradeoff between communications and sensing performances under different transceiver architectures. The achievable rate is computed as $\log_2(1+\text{SNR})$, and the EE is obtained based on the power consumption model in \cite{chen2025energy}, with details omitted due to space limitations. We compare the proposed tri-HBF to fully digital (FD) and HBF architectures. In FD, each antenna has a dedicated RF chain, while in HBF, antennas connect to independent phase shifters that share one RF chain. Owing to the reduced inter-element spacing of DMA ($\lambda_c/5$ vs.~$\lambda_c/2$), we consider these two baseline architectures in two configurations: (i) same number of antennas (SN) and (ii) same antenna aperture with fewer radiating elements (SA).

In Fig.~\ref{fig:AR_vs_SP}, we show the tradeoff between the communications achievable rate and sensing power of the considered schemes. As expected,  the achievable communications rate decreases when sensing power increases. It is observed that the tri-HBF architecture exhibits a strictly smaller tradeoff region than FD SN and HBF SN, mainly due to the physical limitations of DMA, including waveguide loss, beamforming constraints, and the subarray structure. However, 
tri-HBF attains higher communications rates owing to its larger number of radiating elements, which provide a greater spatial gain than the FD SA and HBF SA architectures. In contrast, FD SA and HBF SA achieve higher sensing power, since this metric primarily reflects beam alignment quality, i.e., how well the transmit beam aligns with the steering vector. Fig.~\ref{fig:EE_vs_SP} further presents the tradeoff between communications EE and sensing power. It is clear that the tri-HBF architecture outperforms the compared benchmarks in EE. This is due to the low-power consumption of DMAs and is aligned well with the findings in \cite{castellanos2025embracing, heath2025tri}.

\begin{figure}[t!]
    \hspace{-0.35cm}
    \subfigure[EE vs. achievable rate.]{%
        \includegraphics[width=0.538\linewidth]{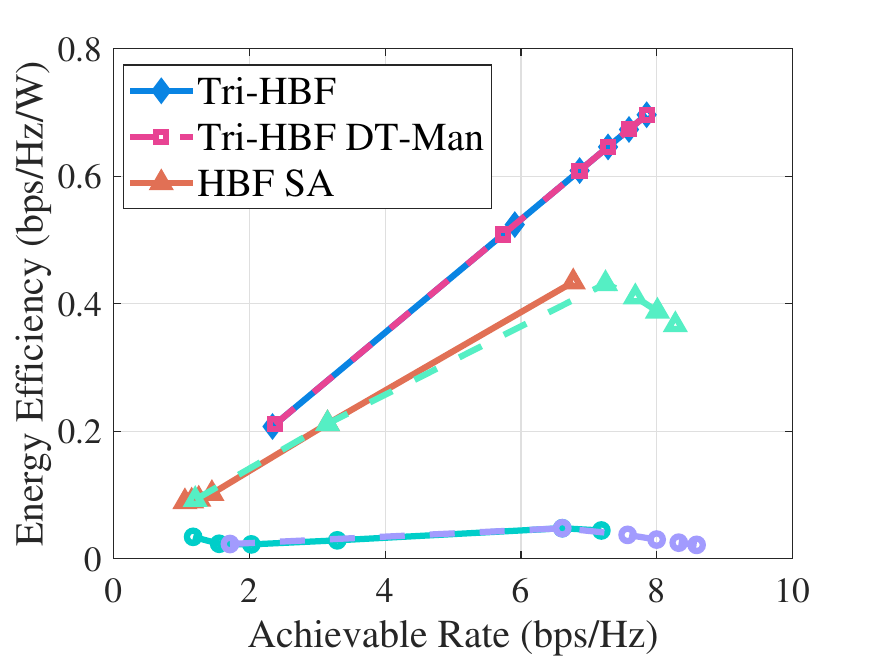}%
        \label{fig:EE_vs_SE}
    }
    \hspace{-0.6cm}
    \subfigure[Sensing power vs. $\Nu$.]{%
        \includegraphics[width=0.538\linewidth]{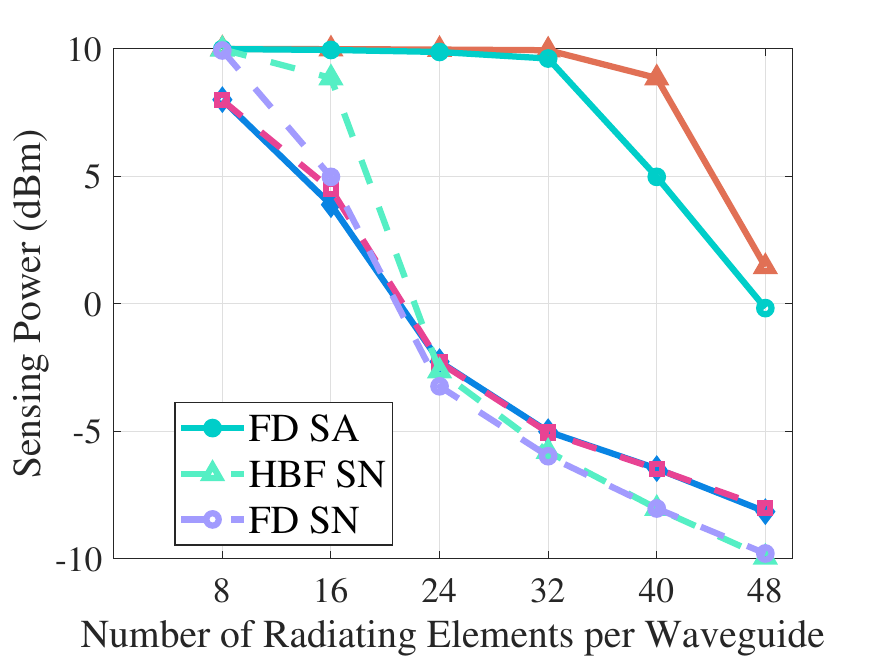}%
        \label{fig:SP_vs_Nu}
    }
        \vspace{-3mm}
    \caption{\small Performance of different transceiver architectures with respect to the number of radiating elements per waveguide $\Nu$.}
    \label{fig:Nu}
\end{figure}

Fig.~\ref{fig:Nu} illustrates the performance comparison of different architectures for $\Nu \in \{8,16,\ldots,48\}$. Following Fig.~\ref{fig:AR_vs_SP}, we fix the weight as $\delta_\mathrm{c} = 0.075$ to balance the performance of communications and sensing functionalities. Fig.~\ref{fig:EE_vs_SE} presents the tradeoff between EE and achievable rate. It can be observed that the proposed tri-HBF architecture achieves substantially higher EE compared with HBF and FD, particularly in the extremely large-scale regime. As the achievable rate increases, the EE of tri-HBF improves linearly. This is because the achievable rate grows significantly while the additional elements do not introduce extra hardware power consumption. What is more, for HBF and FD, the EE first increases notably and then gradually decreases. Fig.~\ref{fig:SP_vs_Nu} further depicts the sensing power versus $\Nu$. In general, the sensing power decreases for all architectures as $\Nu$ increases. This is because the improved channel gain of communications links strengthens the SNR, which dominates the objective function in beamforming design.


\vspace{-0.35cm}
\section{Conclusion}\vspace{-0.25cm}
\label{sec:prior}
We proposed a tri-HBF architecture for mmWave ISAC system with a large antenna array, where digital, analog, and DMA-based beamformers are jointly optimized. By formulating a weighted sum of communications SNR and sensing power, we developed an efficient iterative algorithm with closed-form beamformer updates. Simulation results show the proposed design achieves low computational complexity, fast runtime, stable convergence, and significantly improved communications EE and sensing power compared to conventional architectures, while maintaining competitive communications performance. Future work will extend this design to multi-user, multi-target scenarios, aiming to maximize system-wide sum-rate and sensing accuracy, while addressing practical challenges such as precise channel estimation and inter-user interference management.

\clearpage
    \vspace{-1cm}
\let\OLDthebibliography\thebibliography
\renewcommand\thebibliography[1]{
    \OLDthebibliography{#1}
    \setlength{\parskip}{0pt}
    \setlength{\itemsep}{0pt plus 0.0ex}
}
\section{Acknowledgments}
This work was supported by the Research Council of Finland through the 6G Flagship Program (Grant No. 369116),  
project DIRECTION (Grant No. 354901), project DYNAMICS (Grant No. 367702), and project S6GRAN (Grant No. 370561);  
supported in part by CHIST-ERA through the project PASSIONATE (Grant No. 359817);  
and in part by the HORIZON-JU-SNS-2023 project INSTINCT.

\footnotesize
\bibliographystyle{IEEEbib}
\bibliography{strings,refs}

\end{document}